\documentclass[useAMS,usenatbib]{mn2e}
\usepackage{graphicx}
\usepackage{txfonts}
\usepackage{pstricks}
\usepackage{eepic}
\usepackage{color}
\usepackage[colorlinks]{hyperref}

\newcommand{\figref}[1]{figure (\ref{fig:#1})}
\newcommand{\tblref}[1]{table (\ref{tbl:#1})}
\newcommand{\eqref}[1]{equation (\ref{eq:#1})}
\newcommand{\C}{{\tt C}}
\newcommand{\Cp}{{\tt C++}}
\title[Markov-Chain Monte-Carlo]{Markov-Chain Monte-Carlo\\ A Bayesian Approach to Statistical Mechanics}
\author[T. A. Ottosen]{Thomas Amby Ottosen$^{1}$\thanks{E-mail:
tao05@phys.au.dk}\\
$^{1}$Department of Physics and Astronomy, Aarhus University, Ny Munkegade 120, DK-8000 Aarhus C, Denmark}
\begin{document}

\date{Handed in: 2010 July 5}

\pagerange{\pageref{firstpage}--\pageref{lastpage}} \pubyear{2010}

\maketitle

\label{firstpage}

\begin{abstract}
Since the middle of the 1940's scientists have used Monte Carlo (MC) 
simulations to obtain information about physical processes. This 
has proved a accurate and and reliable method to obtain this information. 
Through out resent years researchers has begone to use the slightly newer 
Markov Chain Monte Carlo (MCMC) simulation. This differs from the ordinary 
MC by using the Markov Chain. MCMC originates from Bayesian statistics.
This method has given researchers a completely new tool to learn something 
about physical systems. One of the fields where MCMC is a good new tool, is
astrophysics. Today MCMC is widely used in simulating power spectra for 
asteroseismic data. Hereby providing the scientists with important new information
of stellar interiors. From our results we see that MCMC delivers a robust and reliable
result with good error estimation. We also learn that MCMC is a power full tool which 
can be applied to a large verity of problems. 
\end{abstract}

\begin{keywords}
Computer science -- Stellar simulation -- Markov-Chain Monte-Carlo
\end{keywords}

\section[Introduction]{Introduction}
In this paper we want to investigate the properties of statistical simulation. We will work on a set of data that we know can be described by a straight line. Since we know how we can describe the data, we also know that we can test various other methods of getting about the fitting of a straight line. The other methods we will compare it with is the least squares fit both weighted an non-weighted. The statistical algorithm we will apply to try and determine the parameters in the straight line fit is called \emph{Markov-Chain Monte-Carlo} (henceforth MCMC). Which has become increasingly popular during resent years. 

Throughout this paper we will go through various things in section 2 we will treat random numbers and problems with different random number generators. We will also address the problem of transforming a uniform random  sample into a normal distributed random sample. In section 3 we will present the data set that are used as the test case for the MCMC. In section 4 we go into details with least squares fitting both with and with out weighting of the data points we will also look at the results when fitting straight lines to our data. Section 5 contains an introduction to Bayesian statistics, and presents the fundamental Bayes' theorem. In Section 6 we go into detail with different types of samplers, we look at two who build on the same principle, these are the Gibbs sampler and the Metropolis-Hastings Algorithm. We also show the result of our Markov-Chain sampling, which give a fairly good representation of our parameter space. Section 7 presents the underlying theory of the Markov-Chain and the Monte-Carlo and finally combine the two into the Markov-Chain Monte-Carlo simulation. Section 8 discuss the results of our MCMC and presents the estimates of $m$ and $b$ for the straight line given in \eqref{line}. Section 9 present some of the perspectives of the MCMC simulations in astrophysics. Since this is becoming more and more used in this field of research. 


\section[Random Numbers]{Random Number Generation and Transformation}
In \cite{NR} a few rules about the random sampling is mentioned that is worth mentioning here as well. Some of the aspects of random number generators one should be aware of are:
\begin{itemize}
	\item Never use a generator based on a \emph{linear congruential generator} (LCG) or a \emph{multiplicative linear congruential generator} (MLCG).
	
	\item Never use a generator with a period less than $\sim 2^{64} \approx 2\times 10^{19}$, or any generator whose period is undisclosed.
	
	\item Never use a generator that warns against using its low-order bits as being completely random. That was once a good advice, but now it seams to indicate an obsolete algorithm (normally a LCG).
	
	\item Never ever use the build in generator of the languages \C ~and \Cp.  Especially {\tt rand} and {\tt srand}, they have no standard implementation and are often badly flawed.
\end{itemize}
Another thing we should be aware of when dealing with random numbers are the over engineered generators which seams to waste resources. 

\begin{itemize}
	\item Avoid generator that take more than $\sim$ two dozen arithmetic or logical operations to generate a 64-bit integer or double precision floating result.
	
	\item Likewise one should avoid the use of generators which is designed for cryptographic use.
	
	\item There are no real need to use generators with periods $> 10^{100}$.
\end{itemize}

From all of this we should learn that: An acceptable random generator must combine at least two (ideally and unrelated) methods. The methods combined should evolve independently and share no state. The combination should be by simple operations that do not produce results less random than their operands, \citep[p. 342]{NR}. 

\begin{figure}
	\input{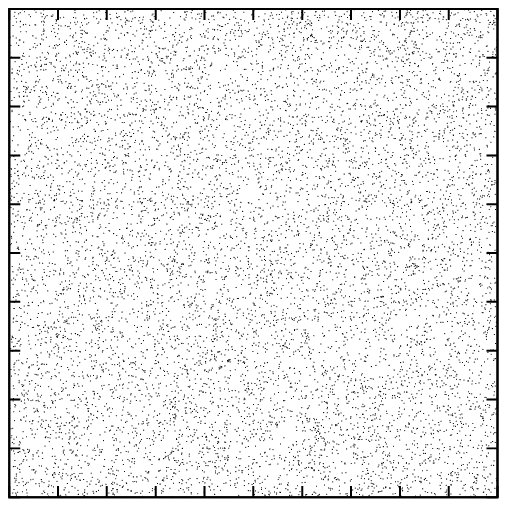}
	\caption{10\ 000 random generated points from the build in {\tt rand} generator.}
	\label{fig:rand}
\end{figure}

\begin{figure}
	\input{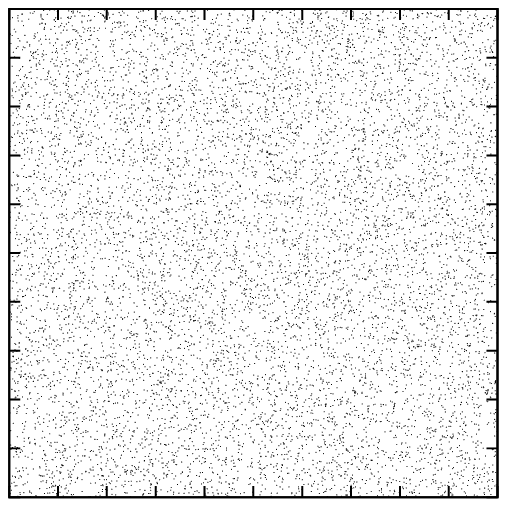}
	\caption{Again 10\ 000 random sampled numbers using a generator described in \citep[p. 342]{NR}}
	\label{fig:NRran}
\end{figure}

If we take a look at the random numbers that are produced by the random number generator {\tt rand} which are the standard one in \Cp ~we find data similar to these ones, see \figref{rand}.
If we now compare \figref{rand} with the one showing the same amount of random sampled points using the random sampler build from the principles from \cite{NR}, \figref{NRran} we see that there are not that much difference in the visual appearance after only 10\ 000 sampled points. But there seams to be a little more clustering of points in \figref{rand}

For this reason we will use the random generator from \cite{NR}. Another thing we will need is a random sampler that can sample a Gaussian distribution, for this purpose we will turn to Box-Muller principle.

\subsection[Box-Muller]{Box-Muller Transformation}
\begin{figure}
	\input{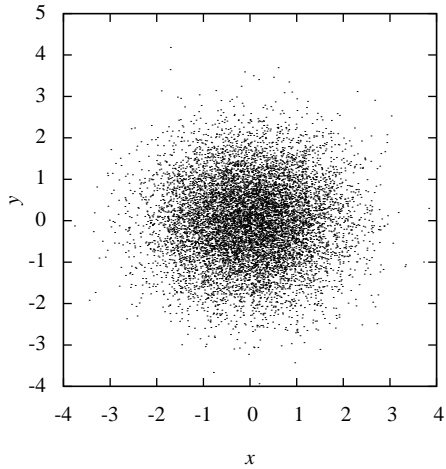}
	\caption{Our 10\ 000 normal distributed points, be aware that the axis have changed if one compares to \figref{rand} and \figref{NRran}.}
	\label{fig:normalBM}
\end{figure}

\begin{figure}
	\input{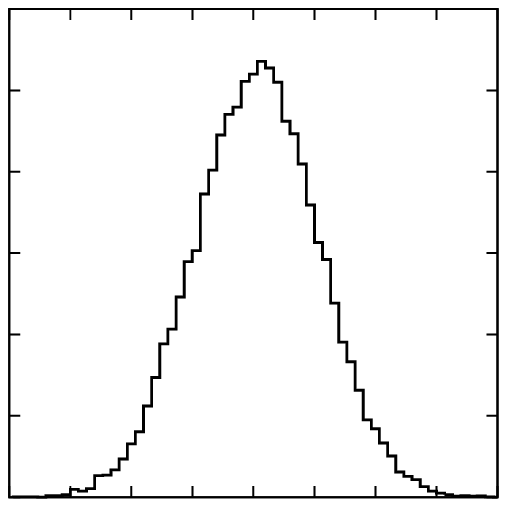}
	\caption{Histogram showing the data from \figref{normalBM}.}
	\label{fig:norm-hist}
\end{figure}

This transformation can be generalised to more than one dimension. Say $x_{1},~x_{2},~\ldots$ are random sampled numbers, with a \emph{joint} probability distribution $p(x_{1},~x_{2},~\ldots) dx_{1}dx_{2}~\ldots$ and if $y_{1},~y_{2},~\ldots$ are functions of the $x$'s (one $y$ per $x$). Then we can describe the joint probability distribution of the $y$'s as:
\begin{equation}\label{eq:normal}
	p(y_{1},~y_{2},~\ldots)dy_{1}dy_{2}~\ldots = p(x_{1},~x_{2},~\ldots) \left|\frac{\partial (x_{1},~x_{2},~\ldots) }{\partial( y_{1},~y_{2},~\ldots)}\right| dy_{1}dy_{2}~\ldots
\end{equation}
where $|\partial (x_{1},~x_{2},~\ldots) /\partial (y_{1},~y_{2},~\ldots)|$ is the Jacobian determinant of the $x$'s with respect to the $y$'s. This principle can be used to transform our uniformly distributed random samples into random samples that are normal (Gaussian) distributed. This distribution is described by:
\begin{equation}\label{eq:gauss}
	p(y)dy = \frac{1}{\sqrt{2\pi}}\exp(-y^{2}/2)dy
\end{equation}
We can now consider the transformation between two uniform deviates on (0,1), $x_{1},~x_{2}$, and two quantities $y_{1},~y_{2}$ as:
\begin{eqnarray}
	y_{1} &=& \sqrt{-2\ln x_{1}}\cos 2\pi x_{2}\\
	y_{2} &=& \sqrt{-2\ln x_{1}}\sin 2\pi x_{2}
\end{eqnarray}
We can now isolate $x_{1}$ and $x_{2}$ by doing this we find that:
\begin{eqnarray}
	x_{1} &=& \exp\left[-\frac{1}{2}\left(y_{1}^{2}+y_{2}^{2}\right)\right]\\
	x_{2} &=& \frac{1}{2\pi} \arctan \frac{y_{2}}{y_{1}}
\end{eqnarray}
Now it is fairly straight forward to calculate the Jacobian determinant  an it turns out we get:
\begin{eqnarray*}
	\frac{\partial(x_{1},x_{2})}{\partial(y_{1},y_{2})} &=& \left|\begin{array}{c c}\frac{\partial x_{1}}{\partial y_{1}} & \frac{\partial x_{1}}{\partial y_{2}}\\ \frac{\partial x_{2}}{\partial y_{1}} & \frac{\partial x_{2}}{\partial y_{2}} \end{array}\right|\\ &=& -\left[\frac{1}{\sqrt{2\pi}}\exp(-y_{1}^{2}/2)\right] \left[\frac{1}{\sqrt{2\pi}}\exp(-y_{2}^{2}/2)\right]
\end{eqnarray*}
From this we see that both $y_{1}$ and $y_{2}$ are distributed according to the normal distribution given in \eqref{gauss}. The result of out normal transformation can be seen in \figref{normalBM}.
In order to convince the suspicious reader that we in fact have created a normal distributed sample we can create the histogram of the data. This can be seen on \figref{norm-hist}. 
From \figref{norm-hist} it is quite obvious that we have created a normal distributed sample since it is nicely Gaussian.

\section[The dataset]{The Data}
\begin{figure}
	\input{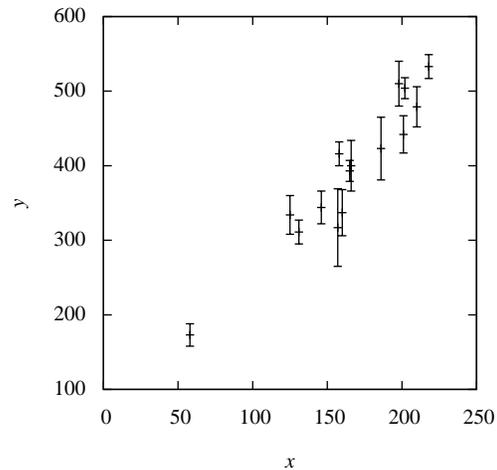}
	\caption{The data chosen for this test of MCMC. We have chosen to weight the data differently in order to see if this have an effect or not.}
	\label{fig:test}
\end{figure}

For the rest of the project we have chosen to work on a set of data, that are made up, but could be data from any part of the physical domain. The data set is presented in \tblref{data}, and an illustration of the data can be seen in \figref{test}. From this figure it is obvious that the data can be described by a straight line given by:
\begin{equation}\label{eq:line}
y = m\cdot x +b
\end{equation}
Since the data can be described by \eqref{line} we know that we can use the linear least squares (henceforth lsq) methods for fitting the optimal straight line, and thereby test the result of our MCMC solution. 
\begin{table}
	\centering
	\caption{The data used in the MCMC analysis}
	\label{tbl:data}
	\begin{tabular}{@{} c c c}
\hline
x & y & dy\\
\hline
58&173&15\\
125&334&26\\
131&311&16\\
146&344&22\\
157&317&52\\
158&416&16\\
160&337&31\\
165&393&14\\
166&400&34\\
186&423&42\\
198&510&30\\
201&442&25\\
202&504&14\\
210&479&27\\
218&533&16\\
\hline
\end{tabular}

\end{table}

We are planing to do the lsq in two versions both and ordinary non-weighted lsq (hence OLS) and a weighted lsq (henceforth WLS). For computing the lsq we use the least squares routine developed during the course.

Another thing one should notice about the data is that they are weighted, see \figref{test}. We also notice that not all the data points are weighted equally, the reason for this is to try and show that there is also a significant difference between the results of OLS, and WLS.

\section[Least Squares]{Least Squares Fitting}
The point of lsq is to determine $m$ and $b$ of \eqref{line}. This could be done for any arbitrary set of data, not only data which falls on a straight line. The only requirement for our routine is that it is a linear set of data. Now would be a good time to fit a line to data so we have some idea about the parameters. 

\subsection[Non-Weighted Least Squares]{Ordinary Linear Leas Squares Fitting}
\begin{figure}
	\input{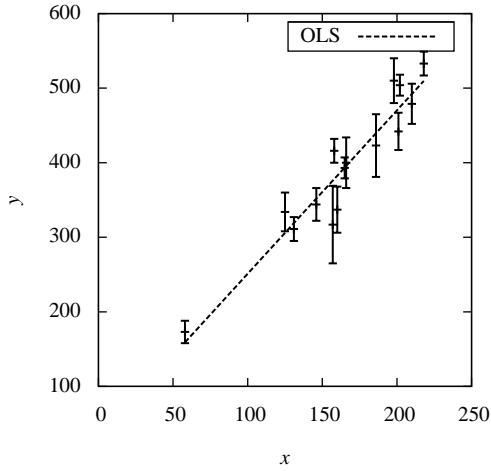}
	\caption{Here is the non-weighted OLS fit, where $m = 2.19$ and $b = 32.00$}
	\label{fig:OLS}
\end{figure}

We will start with the OLS. The OLS routine is included in the {\tt lsq.cpp} program which is simply a modified version of what we have already developed in the \emph{linear least squares} exercise. The modification is simply that we have removed the weighting of the individual data points. The OLS solution is $m = 2.19$ and $b = 32.00$. The result of this fit can be seen in \figref{OLS}. This seam at first sight to be a rather good fit, but since we neglected the weights of the individual data points. This is not the optimal fit. Had the data been equally weighted e.g. if we did not have any information about the weighting of individual data points. 

\subsection[Weighted Least Squares]{Weighted Linear Least Squares Fitting}
\begin{figure}
	\input{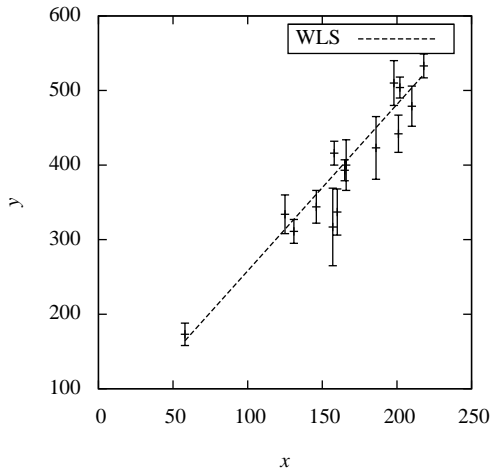}
	\caption{Weighted LSQ fit, with $m = 2.23$ and $b = 34.80$, to the data set.}
	\label{fig:WLS}
\end{figure}

The other approach of the is the WLS is a solution to the data. Since not all the data points are not equally weighted. The WLS takes the individual weights into account. The WLS fit to the data gives us the following values for $m = 2.23$ and $b = 34.80$.  From \tblref{lsq-fit} we can see that we get somewhat higher values of $m$ and $b$ than that of the OLS.  If we plot the result of the WLS, we get the following result, see \figref{WLS}.

\begin{figure}
	\input{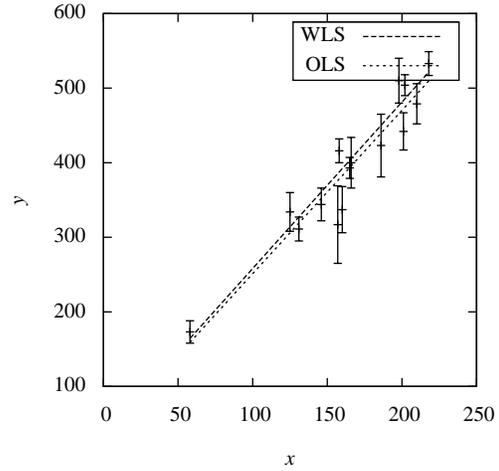}
	\caption{comparison between the OLS and the WLS lines}
	\label{fig:COMP}
\end{figure}

When comparing the two plots \figref{OLS} and \figref{WLS} it is not obvious that there is any difference at all. 
But if the two results are plotted in the same plot we see a clear difference see \figref{COMP}. 
\begin{table}
\caption{This table shows the results of the lsq fitting to the data for the two methods of lsq used here.}
\label{tbl:lsq-fit}
\centering
\begin{tabular}{@{}l r@{.}l r@{.}l r@{.}l}
\hline
 & \multicolumn{2}{c}{{\bf OLS}}& \multicolumn{2}{c}{{\bf WLS}} & \multicolumn{2}{c}{{\bf Diff.}}\\
\hline
$m$ & 2&19 & 2&23 & 0&04\\
$b$ & 32&00 & 34&80 & 2&80\\
\hline
\end{tabular}
\end{table}
We can now see that our expectation that the WLS is a better solution of the line that describe the data, than that of OLS.
Since we now have the optimal solution for the straight line we can now start our statistical modelling of the data. For this we are going to use MCMC.

\section[Bayes' Theorem]{Bayes' Theorem and the link to MCMC}
In statistics there have been two branches, competing about which one was the correct one to use. Scientists have been changing from one to the other and back again as time has gone. The two branches are known as ``\emph{frequentists}'' and ``\emph{Bayesians}''. Nowadays the Bayesian statistics are the dominant branch. It is also from this branch that the MCMC have rose. 

Bayesian statistics generally builds on the theory of Rev. Thomas Bayes (1702 -- 1761). The reason for his importance is that he formulated the fundamental theorem now known as Bayes' theorem.
Bayes theorem generally shows the relation between one conditional probability and its inverse, e.g. the probability of a hypothesis given observed evidence and the probability of that evidence given the hypothesis. In other words; \emph{Data changes probability}.

In a little more strict formulation Bayes' theorem states:
\begin{quote}
	The posterior probability (i.e. after evidence $D$ is observed) of a hypothesis $H$ in terms of the prior probabilities of $H$ and $D$, and the probability of $D$ given $H$. It implies that evidence has a stronger confirming effect if it was more unlikely before being observed.
\end{quote}
 If we want to write Bayes' theorem as an equation we can do it in the following manor.

Bayes theorem relates the conditional and marginal probabilities of the events $A$ and $B$, where $B$ has a non vanishing probability:
\begin{equation}\label{eq:bayes}
	p(\mathbf{A}|\mathbf{B}) = \frac{p(\mathbf{B}|\mathbf{A})p(\bmath{A})}{p(\bmath{B})}
\end{equation}
In this equation $p(\bmath{A)}$ is the marginal probability or \emph{prior} probability (henceforth prior). It is a prior in the sense that it does not take any information about $\bmath{B}$ into account. $p(\mathbf{A}|\mathbf{B})$ in the conditional probability of $\bmath{A}$, given $\bmath{B}$. this also called the posterior probability (hence posterior) since it depends on the specific value of $\bmath{B}$. $p(\mathbf{B}|\mathbf{A}) $ is the conditional probability of $\bmath{B}$ given $\bmath{A}$, this is also called likelihood which will be the term used hence forth. The \eqref{bayes} is normalized by the prior of $\bmath{B}$. 

\section[Metropolis]{Sampler}
In order to make our MCMC work we need to have a control if a new sampled point should be accepted or not. There are several ways of doing that one of these ways are known as the \emph{Metropolis-Hastings} algorithm, \cite{Hast}, and the \emph{Gibbs} sampler, \cite{CG,CJ} .  We will go through the theory of both of them, but in our MCMC we will use the Metropolis-Hastings algorithm.

\subsection[Gibbs Sampling]{The Gibbs Sampler}
The Gibbs sampler is an algorithm used for generating a sequence of samples from the joint probability distribution of two or more random variables, see \eqref{normal}. The purpose a sequence like this is to approximate the joint distribution, or compute an integral, e.g. an expectation value. In principle the Gibbs sampling is a special case of the of the Metropolis-Hastings algorithm. Since we are not going to apply the Gibbs sampling we will not go into further detail with here, more information on the Gibbs sampler can be found in \cite{Gey, NR,CJ}.

\subsection[Metropolis-Hastings]{The Metropolis-Hastings Algorithm}
The Metropolis-Hastings algorithm is the heart of our MCMC program. Since it is a method for obtaining a sequence of random numbers from a probability distribution for which direct sampling is difficult, or impossible. The sampled sequence can then be used for approximating the distribution, e.g. by producing a histogram. The algorithm can be presented in the following way:

Unless a transition probability function$p(\bmath{x}_{2}|\bmath{x}_{1})$  that satisfy the detailed balance equation
\begin{equation}
	\pi(\bmath{x}_{1}) p(\bmath{x}_{2}|\bmath{x}_{1}) = \pi(\bmath{x}_{2})p(\bmath{x}_{1}|\bmath{x}_{2})
\end{equation}
There is no possible way of continuing. That is when we need the Metropolis-Hastings algorithm. Which say that we should pick a \emph{proposal distribution} $q(\bmath{x}_{2}|\bmath{x}_{1})$.  This distribution can be what ever you like, as long at the succession of the steps generated reach everywhere in the region of interest. One example could be that $q(\bmath{x}_{2}|\bmath{x}_{1}) $ might be a multivariate normal distribution centred around $\bmath{x}_{1}$. 

Now in order to generate a step starting at $\bmath{x}_{1}$, we first generate a \emph{candidate point} $\bmath{x}_{2c}$ by drawing from the proposal distribution. The second step is to calculate the \emph{acceptance probability}  $\alpha(\bmath{x}_{1},\bmath{x}_{2c})$ this can be done using the following equation
\begin{equation}
	\alpha(\bmath{x}_{1},\bmath{x}_{2c}) = \min \left(1,\frac{\pi(\bmath{x}_{2c})q(\bmath{x}_{1}|\bmath{x}_{2c})}{\pi(\bmath{x}_{1})q(\bmath{x}_{2c}|\bmath{x}_{1})}\right)
\end{equation}
The final and third step is with the probability $\alpha(\bmath{x}_{1},\bmath{x}_{2})$ to accept the candidate point and set $\bmath{x}_{2} = \bmath{x}_{2c}$. if this is not possible the point should be rejected and hereby leave the point unchanged meaning $\bmath{x}_{2} = \bmath{x}_{1}$. The netto result of this process is a transition probability, \cite{NR,Hast,CG,CJ}.
\begin{equation}
	p(\bmath{x}_{2}|\bmath{x}_{1}) = q(\bmath{x}_{2}|\bmath{x}_{1})\alpha(\bmath{x}_{1},\bmath{x}_{2})\qquad (\bmath{x}_{2} \neq \bmath{x}_{1})
\end{equation}

\subsection[Sampled Results]{Results of our Metropolis-Hastings Algorithm}
\begin{figure}
	\input{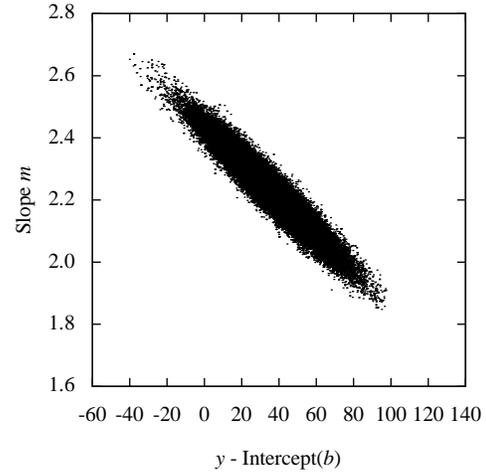}
	\caption{The sampled data points used to make our description of our parameter space.}
	\label{fig:SAMP}
\end{figure}

In \figref{SAMP} we can see the sampled data plotted in two dimensions. This mean that we now get a \emph{direct} impression of our parameter space. Since the density of the dots in \figref{SAMP} is an expression of the number of occurrences. So just from this simple plot we can already estimate that  the parameter $m$ should be around $m \in [2.0;2.4]$ and $b$ is then from this visual inspection determined to be $b \in [0;70]$.
Of cause these are very rough estimates. They still tells us that both the OLS and the WLS seam to be fairly good estimates. Let us move on to the MCMC simulation.

\section[MCMC]{The Markov-Chain Monte-Carlo}
Let us begin this section by breaking down the term MCMC into the points of which it is build. So MCMC becomes MC---MC and we will then look at what a Markov-Chain is:

\subsection[Markov-Chain]{Markov-Chain}
The Markov-Chain is a discrete random process with the property that the next state depends only in the current state. It is named after Andrei Andreyevich Markov (1856 -- 1922). The Markov-Chain is a mathematical tool used for statistical modelling, it can be thought of as a frog jumping among several lily-pads, where the frog's memory is short enough that it does not recall what lily-pad it was last on. and therefore the next jump will only be influenced by where it is now.

A little formally we can state what the what the Markov-Chain is all about like this:

A Markov-Chain is a discrete random process with Markov property that goes on forever. A discrete random process is one where a system is in a certain state at each \emph{step}, with the state changing randomly with the steps. The steps are often thought of as time, but they can as well refer to physical distance or any other discrete measurement. The steps are just the natural numbers, and the the random process is a sort of mapping of these to states of the system. The Markov property states that the conditional probability distribution for the system,  at the next step given the current state depends only on the current state of the system, and not on the state of the system at any previous steps.
\begin{equation}
	p(\bmath{x}_{n+1}|\bmath{x}_{1},\bmath{x}_{2},\ldots ,\bmath{x}_{n}) = p(\bmath{x}_{n+1}|\bmath{x}_{n})
\end{equation}
Since the system changes randomly, it is generally impossible to predict the exact state of the system in the future. However, the statistical properties of the system's future can be predicted. In many application it is these statistical properties that are important, \cite{Sag,Gey,CHS}.

\subsection[Monte-Carlo]{Monte-Carlo}
Since the 1940's when scientists at Los Alamos Laboratories in the US invented the Monte-Carlo method (hence MC). Since then MC has been widely used in computational physics. Mainly due to it's computational friendliness, \cite{MC}. MC or MC integration, as it is often called is a method used for obtaining knowledge about systems which are beyond the reach of analytical methods \cite {MC}.
Suppose that we randomly pick $N$ uniformly distributed points. in a multidimensional volume $V$. These points will be referred to as $x_{0}, ~x_{1}, ~\ldots , ~x_{N-1}$. Then we apply the underlying theorem of MC integration. 
\begin{equation}\label{eq:MCb}
	\int f dV \approx V \langle f \rangle \pm V \sqrt{\frac{\langle f^{2}\rangle-\langle f \rangle^{2}}{N}}
\end{equation} 
This theorem seen in \eqref{MCb} estimates the integral of the function $f$ over the multidimensional volume.  The angle brackets in \eqref{MCb} is simply the arithmetic mean offer the $N$ sample points,
\begin{equation}\label{eq:MCf}
	\langle f\rangle \equiv \frac{1}{N}\sum_{i = 0}^{N-1}f(x_{i})\qquad\quad \langle f^{2} \rangle \equiv \frac{1}{N}\sum_{i = 0}^{N-1}f^{2}(x_{i})
\end{equation} 
One should notice the plus-minus sign in \eqref{MCb}, this is only an one standard deviation error estimate for the integral, not a rigorous bound. 

The MC approach have been used very extensively in the past, and through this approach we have learnt a lot, and it has proved itself, as a robust way of determine integrals numerically. So if this method is so good an robust why do we then need a new one? 

There are several reasons, first of all most MC simulations  uses only a pseudo-random sampling. Then depending on the random sampler one uses it will be determined whether or not it is a good estimate. Since it is dependant on the kind of random sampler one uses for the MC. 

There is no single Monte-Carlo method; instead the therm MC describes a large and widely-used class of approaches. However, these approaches tend to follow a particular pattern:
\begin{enumerate}
	\item one has to define a domain of possible inputs.. 
	\item then it is time to generate some inputs randomly from the domain using a certain specified probability distribution. 
	\item one needs to perform a deterministic computation using the inputs.
	\item one needs to combine the results of the individual computations into the final result.
\end{enumerate}

We will not go into any further detail with the underlying theory of the Monte-Carlo method, since this has been the subject of one of the exercises in the course. In the mid 1980's a new approach to MC was presented to the world this was known as MCMC, the main difference between MC and MCMC is that MC is based on a random sampling while MCMC builds on the principle of random walk. 

\subsection[Markov-Chain Monte-Carlo]{Markov-Chain Monte-Carlo}
Like the MC integration, MCMC is a random sampling method, but unlike the MC integration the goal of the MCMC is not to sample a multidimensional region uniformly. The goal is rather to visit a point $\bmath{x}$ with a probability proportional to a given distribution function $\pi(\bmath{x})$. The distribution $\pi(\bmath{x})$ is not quite a probability, because it might not be normalized to unity when integrated over the sampled region. However it is proportional to the probability.

The reason for applying this is that MCMC provide a powerful way of estimating parameters of a model and their degree of uncertainty. The typical case is that we have a set of data $\bmath{D}$, and we are able to calculate the probability of the dataset \emph{given} the values of the parameters of the model $\bmath{x}$, which yield $P(\bmath{D}|\bmath{x})$. if we assume the prior $P(\bmath{x})$ Bayes' theorem, \eqref{bayes}, will tell us that $\pi(\bmath{x}) \equiv P(\bmath{D}|\bmath{x})P(\bmath{x})$. In the form of Bayes' theorem given in \eqref{bayes}, the normalization is known, but it might be unknown. If we assume that the normilaztion is unknown,then $\pi(\bmath{x})$ is will not be the normalized probability density. But if we can sample from it, we can achieve information about any quantity of interest. This is interesting since we can use this fact to obtain information on the normalized probability density simply by observing how often we sample a given volume $d\bmath{x}$. Another maybe even more useful property is that we can observe the distribution of any single component or set of components. So one of the questions we should ask is; ``Why is there at all a need to use MCMC instead of MC?'' This is an interesting question, but it is fairly easy to answer, one of the major advantages of MCMC is that it automatically puts the samples points preferentially where $\pi(\bmath{x})$ is large. In a high-dimensional space or where $\pi(\bmath{x})$ is expensive to compute, this can be advantageous by several orders of magnitude.

Our next step is to assemble both parts of our Markov-Chain using the Metropolis-Hastings algorithm and our Monte-Carlo algorithm in order to get a fully working Markov-Chain Monte-Carlo  \cite{CHS,Gey,Sag,NR,Cap}.

\section[Results]{Results}
We now need to extract some results from our simulation. This is done using the sampled data, from these we can create a histogram in either direction, of the parameter space. So we will in the end have a histogram for the parameter $m$ and one for the parameter $b$. The way we go about making these histogram is by using the routine {\tt gsl-histogram} from the GNU scientific library.  This routine is very simple to use since it is just passed on the commandline as:

{\tt \$ ~$\rangle$ cat file.dat | gsl-histogram $x_{\min}$ $x_{\max}$ $n_{\mathrm{bins}}$}

we can pipe the output directly into graph or to a new data file for plotting with gnuplot. Even for rather large data sets as the one we presents with 100\ 000 data points we still get our result from the gsl routine almost instantly. This ensures a high efficiency. However it also presents a problem. Since the histogram data are created by running the makefile as 

{\tt \$ ~$\rangle$ make histogram}

we need to compile the makefile several times to get it all done. Therefore I have chosen to make a shell-script to ensure that everything is done in the right order this script is called {\tt domake.sh}.
Let us have a look at the results.

\subsection[Estimating $m$]{Estimating $m$}
\begin{figure}
	\input{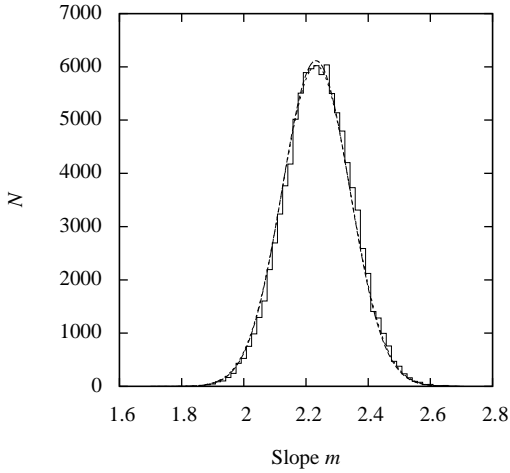}
	\caption{Estimation of the parameter $m$.}
	\label{fig:ESTM}
\end{figure}

We now have our sampled data from our MCMC and therefore want to determine $m$. We can estimate $m$ by making a histogram along the $m$-axis of our sampling, see \figref{SAMP}. This will result in the histogram showed in \figref{ESTM}. In order to determine where the max is we can fit a Gaussian to this histogram. This can be done in two ways. Either by simple algebraic solving the equation of the Gaussian for the values provided by our sampling. This is the line with the long dashes in \figref{ESTM}.  The short dashed line indicate a Gaussian fit by the MCMC values. 

\begin{figure}
	\input{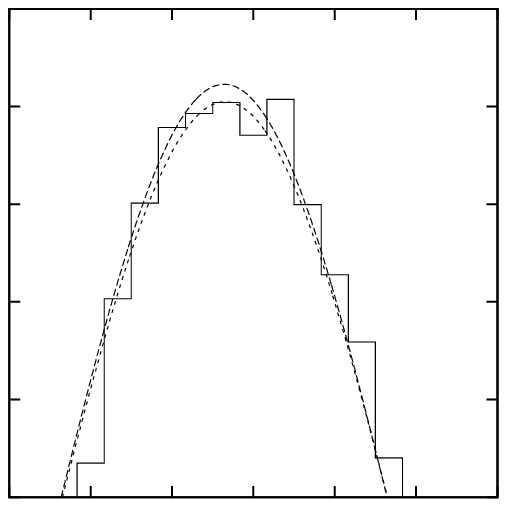}
	\caption{ Zoom in on the upper half of \figref{ESTM}. Estimation of the parameter $m$.}
	\label{fig:ESTm}
\end{figure}

These lines are not easy to separate from each other, which is why in \figref{ESTm} have a zoom of the upper part of \figref{ESTM}. From this zoom one can as well see that the maxima's of the two lines are not exactly above each other. But they are shifted by a very tiny amount.

From the estimation we get the results shown in \tblref{fits}. And we see that the determination of the parameter is far better regarding the uncertainties associated since we here are looking at uncertainties of $\pm 5 \%$ in $m$, for the algebraic fit and the MCMC fit, while we for the WLS fit have an uncertainty of $\pm 15 \%$.

Another thing we should take notice of is that fact that with the MCMC we approximates the final result of the WLS far better than that of the algebraic solution even though it is only .02 percent that separates them.

\subsection[Estimating $b$]{Estimating $b$}
\begin{figure}
	\input{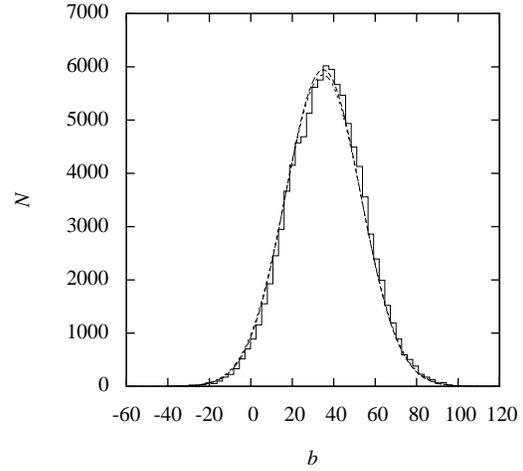}
	\caption{Estimation of the parameter $b$}
	\label{fig:ESTB}
\end{figure}

We now have our sampled data from our MCMC and therefore want to determine $b$. We can estimate $b$ by making a histogram along the $b$-axis of our sampling, see \figref{SAMP}. This will result in the histogram showed in \figref{ESTB}. In order to determine where the max is we can fit a Gaussian to this histogram. This can be done in two ways. Either by simple algebraic solving the equation of the Gaussian for the values provided by our sampling. This is the line with the long dashes in \figref{ESTB}.  The short dashed line indicate a Gaussian fit by the MCMC values. 

\begin{figure}
	\input{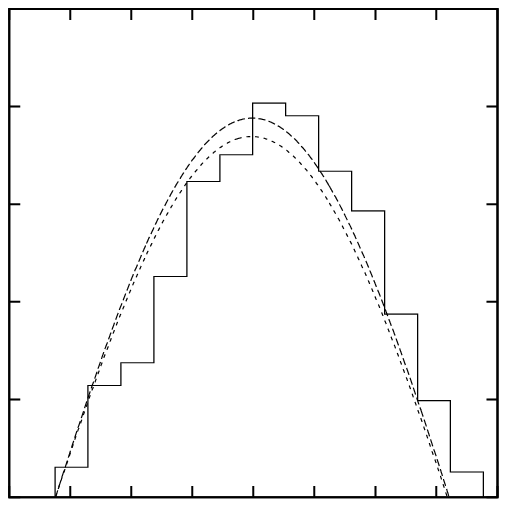}
	\caption{Zoom in on the upper half of \figref{ESTB}. Estimation of the parameter $b$.}
	\label{fig:ESTb}
\end{figure}

These lines are not easy to separate from each other, which is why in \figref{ESTb} have a zoom of the upper part of \figref{ESTB}. From this zoom one can as well see that the maxima's of the two lines are not exactly above each other. But they are shifted by a very tiny amount.

From the estimation we get the results shown in \tblref{fits}. Again we see how the parameters fits the sampled data very well and we also see a nice correspondence between WLS and MCMC fits.

\begin{table*}
	\centering
	\caption{Table showing the results of the algebraic solution versus the MCMC and the lsq fits.}
	\label{tbl:fits}
	\begin{tabular}{@{}c r@{.}l r@{.}l@{ $\pm$ }r@{.}l r@{.}l@{ $\pm$ }r@{.}l r@{.}l@{ $\pm$ }r@{.}l}
	\hline
	 & \multicolumn{2}{c}{{\bf OLS}} & \multicolumn{4}{c}{{\bf WLS}} & \multicolumn{4}{c}{{\bf Algebra}} & \multicolumn{4}{c}{{\bf MCMC}}\\
	 \hline
	 $m$ & 2&1910 & 2&2325&0&3309 & 2&2321&0&1087 & 2&2324&0&1103\\
	 $b$ & 32&0040 & 34&8495&55&2679 & 34&9215&18&1203 & 34&824&18&4226\\
	 \hline
	 \end{tabular}
\end{table*}

\section[Perspectives of MCMC]{Perspectives  and applications of MCMC}
In this section we will briefly discuss some of the issues for which we can apply MCMC in the field of astrophysics. Within astrophysics the MCMC technique have become very popular over resent years. Especially in fields as cosmology and extra-solar planet hunting. So what do the scientists use this power full tool for?

\subsection[Exoplanets]{Extra-solar planet hunting}
In the field of Extra-solar planet (hence exoplanet) hunting MCMC is widely used to determine the planet-star -radius ratio, inclination angle, etc, see \cite{gib09, gib10b} for further details of the exact method used. The systemic inclination is of severe importance an it is therefore necessary to get a very good and robust determination of this value. Even though we know it to be $90^{\circ}\pm 5 - 10^{\circ}$ since we would not see the transit when the planet passes in front of the star otherwise. Likewise it is also favourable that one gets very robust uncertainties with the MCMC approach. 

In general it is possible to derive a lot of different parameters which are of interest in the exoplanet domain by the method of MCMC, see \cite{Marie} whom uses MCMC to derive all systemic parameters and their uncertainties. Other references might be \cite{gib10a, WLCDS}

Another very interesting thing in the astronomy domain is a new website called astrometry.net. This website can calibrate any picture of the night sky simply by using a MCMC algorithm \cite{Lang}. So if you are wondering what you are looking at when looking upon the night sky you can simply take a photo with your cell phone and upload it to the website and you will instantly get information on that region of the sky.

\subsection[Cosmology]{MCMC and cosmology}
Another field where MCMC have been applied is cosmology this has basically been done by the same research team who are responsible for the astrometry.net site. They have undertaken an enormous task by starting to look at the Hipparcos catalogue and try to classify the stars in the sample by their velocity distribution for this they have used MCMC see \cite{BH,BH2,BH3}.

\section[Conclusion]{Conclusion}
During this paper we have gone through some of the basic theory of MCMC and we have tested the principle of MCMC on a rather simple dataset, and see that this is indeed a robust and powerful tool when working with observed data. We can conclude that the MCMC method is just as good as the standard WLS, the major advantage of the MCMC compared to the WLS is that the uncertainties is a factor three better than those of the WLS. Some of the disadvantages is however that the MCMC is rather complex, and depending on your parameter space it might not be straight forward.  Other methods we could have compared could be optimisation by the simplex method. We have chosen not to do that.


\section*{Acknowledgments}

I would like to express my gratitude to Professor Dmitri Fedorov for advice and guidance during the 
programming. I would also like to thank Professor David W. Hogg for the inspiration to write this paper,
likewise Professor Bill Press, for his inspiring lectures on MCMC at the Heidelberg Summer School last year.

\appendix

\bsp

\label{lastpage}

\end{document}